# Hysteretic magnetoresistance in polymeric diodes

Sayani Majumdar
*Department of Physics and Center for Functional Materials, Åbo Akademi University, Turku 20500, Finland.*
*Wihuri Physical Laboratory, Department of Physics and Astronomy, University of Turku, Turku 20014, Finland.*

Himadri S. Majumdar, Harri Aarnio and Ronald Österbacka
*Department of Physics and Center for Functional Materials, Åbo Akademi University, Turku 20500, Finland.*

*Abstract*

We report on hysteretic organic magnetoresistance (OMAR) in polymeric diodes. We found that magnitude and lineshape of OMAR depends strongly on the scan speed of the magnetic field and on the time delay between two successive measurements. The time-dependent OMAR phenomenon is universal for diodes made with various polymers. However, the width and magnitude of OMAR varied with the polymeric material. The suggestive reason for this hysteretic behavior are trapped carriers, which in presence of a magnetic field changes the ferromagnetic ground-state of the polymer leading to long spin relaxation time. These experimental observations are significant for clarification of the OMAR phenomenon.

**Introduction**

Organic magnetoresistance (OMAR) is a recently discovered phenomenon where diodes with organic semiconductors exhibit large change in resistance under a small applied magnetic field (typically 100 – 200 mT) at room temperature [1,2]. Though effect of magnetic field on organics is not a new phenomenon [3], this new discovery has generated tremendous prospect for application as magnetically controlled flexible optoelectronic or sensor devices. The physical nature of charge (spin) dynamics of these devices under magnetic field is an ongoing field of research. Different models, namely, the excitonic model, [4] the triplet-exciton polaron quenching model [5] and the bipolaron model [6] have been proposed for explaining this phenomenon. The bipolaron model proposes that magnetoconductance (MC) curves are either a Lorentzian $B^2/(B_0^2+B^2)$ or a specific non-Lorentzian $B^2/(|B|+B_0)^2$, where B is the applied magnetic field and $B_0$ is the full width at half maximum (FWHM) of the MC curves. This is the only quantitative measure of the magnetic field dependence of OMAR line-shape till date. Although this model successfully described the positive and negative sign of MC together with fitting of the MC line shapes, the widening of MC curve (increased $B_0$) upon changing MC sign from positive to negative with increasing branching ratio (ratio of rate of bipolaron formation and rate of hopping to any environmental site) could not be explained well. Two recent papers further suggested that neither the bipolaron model nor the excitonic models are adequate for explaining other experimental observations. [7,8]

In this letter, we report on the dependence of both the OMAR lineshape and magnitude on different measurement conditions. Positive OMAR was observed in the diodes at higher bias currents, as reported earlier. [7] The OMAR magnitude and line shapes change dramatically based on the magnetic history of the devices and vary considerably with the magnetic field scan speed and the time delay between successive measurements. OMAR magnitude is smaller and the FWHM of the line shape is broader with faster magnetic field scan. These results would help in understanding the true physical picture of the OMAR effect and also call for care during measurement and interpretation of magnetotransport data in organic devices.

**Experimental**

The device structure for the diodes used in the experiment is indium tin oxide (ITO)/ poly(3,4-ethylenedioxythiophene)-poly(styrenesulphonate) (PEDOT:PSS)/ regio regular poly (3-hexyl thiophene) (RRP3HT)/Lithium fluoride (LiF)/Au. The ITO coated glass electrodes were coated with a very thin layer of PEDOT:PSS and annealed at 120 ºC for 15 minutes. The polymer RRP3HT, obtained from Aldrich, was spin-coated from a dicholorobenzene or chloroform solution and annealed at 120 ºC for 15 minutes. Finally the lithium fluoride and/or the top electrode was vacuum evaporated to complete the device structure. The device preparation was done in a nitrogen-filled glove-box and using anhydrous solutions. After fabrication, the devices are transferred via a nitrogen chamber to the cryostat placed in between the pole pieces of the electromagnet capable of producing up to 300 mT magnetic field. The resistance of the device is then measured by sending a constant current through the device and measuring the voltage drop in varying magnetic field in the temperature range 100 – 300 K. For each scan, first magnetic field was stabilized and then current was sent through the sample and corresponding voltage was measured. Magnetic field scan for normal MR measurement was done from 0 to 300 mT and from 0 to – 300 mT while for the hysteresis measurements the magnetic field was scanned from 0 to 300 mT, from 300 to – 300 mT and back to 300 mT with different field sweeping rates.

**Results and Discussions**

Fig. 1(a) shows the scan speed dependence of OMAR at room temperature of a typical diode device with ITO/PEDOT:PSS as the hole injecting and Al as the electron injecting electrode. All the measurements were performed using same bias current of 1 µA and by sweeping the magnetic field from 0 to 150 mT varying the scan speeds $\left(s = \frac{\Delta B}{\Delta t}\right)$, where $\Delta B$ is the variation of magnetic field in time $\Delta t$. The OMAR values for the slower $s$ is concurrent with data previously reported [7,9] for diodes made of RRP3HT from dichlorobenzene. MR is defined as $\%MR = \frac{R_B - R_0}{R_0} \times 100 = \frac{\Delta R_B}{R_0} \times 100$, where $R_B$ is the device resistance under external magnetic field B and $R_0$ is the zero field resistance. For slower $s$, OMAR magnitude is larger and FWHM of the line shape is narrower. With increasing $s$, OMAR magnitude starts decreasing and the FWHM also becomes broader. It was also reported before [9] that the line shapes of OMAR traces under high bias currents are better fits with a $B^{0.5}$ dependence rather than the Lorentzian $B^2/(B_0^2+B^2)$ or non-Lorentzian $B^2/(|B|+B_0)^2$ line shape. The $\sim B^{0.5}$ dependence in the high-field region is found to be retained for all $s$ (Fig. 1(b)). To confirm that this change in resistance only occurs due to the presence of both the magnetic and electric field, the diode resistance was measured for a constant current (1µA) in a changing magnetic field (B = 0 – 150 mT with $s$ = 20 µT/sec) and in absence of magnetic field (B = 0) (Fig. 1(c)) over the same time period. In absence of B the device resistance showed very little change in resistance over a period of 2 hours. However, with increasing B the device resistance changed in the same way as indicated in Fig. 1(a) and (b).

Another important observation is the asymmetry of OMAR line shape for different time delay between consecutive scans ($t_d$) above and below B=0. When the magnetic field is scanned from 0 to +300 mT and immediately scanned from 0 to -300 mT field, the OMAR trace is completely asymmetric on both sides of B=0 as shown in Fig. 2(a). The asymmetry of the OMAR plot across the positive and negative magnetic field scan has not been addressed till date. Depending on $t_d$, the OMAR values and line shapes also change significantly. The symmetric behavior of OMAR lineshapes are only regained when $t_d$ is sufficiently large, as can be seen from the plot the resistance (%MR) becomes symmetric when the $t_d$ is about 10 mins. This, in turn, also effects the interpretation of the magnitude of the OMAR. The %MR at B=0 after the positive scan for different time intervals (inset of Fig. 2(a)) shows the variation more elaborately. This residual %MR has significant effect on the lineshape and the magnitude of %MR in subsequent scans. It is to be noted that during the whole course of the 0 to +300 mT scan, the $t_d$ and then the 0 to -300 mT scan, a constant current of 10 µA is reatined across the device.

To separate the effect of electric and magnetic field on the relaxation process of the device, we measured the device resistance as a function of time after the magnetic field is scanned from 0 to 50 mT and then swiched off (Fig. 2(b)). Simultaneous removal of electric and magnetic field (red line) causes restoration of the pristine device state within 5-7 minutes (as also observed earlier) while only removal of magnetic field does not allow full restoration of the initial device resistance state (complete relaxation) even in hours. This long relaxation in the devices under electric field after removal of magnetic field suggests that in presence of injected carriers the existing magnetic environment of the devices are extended over a longer period of time, effectively changing the spin dynamics of the charge carriers in the device.

To clarify the time-scale in which the sample regains its original resistance state under bias stress, we measured the complete magnetotransport hysteresis loops of the devices with different scan speeds, sweeping the magnetic field from 0 to +300 mT, + 300mT to – 300 mT and – 300 mT to + 300mT, consecutively. Fig. 3 shows the scan result of a typical RRP3HT diode with a constant current of 1 µA with two different magnetic field scan speeds ($s$). For Fig. 3(a), $s$ = 10 mT/sec and fig. 3(b) $s$ = 20 µT/sec. As evident from Fig. 3(a), for faster scan speed, the device resistance does not change with change in magnetic field (after reaching a saturation value) and do not return to its initial value. Whereas for fig. 3(b), the OMAR line shape starts to relax after a long time showing a closed hysteresis loop.

In polymeric samples we have recently observed weak ferromagnetism in the ground state which depends strongly on the morphology of the polymer.[10] The cause for such ferromagnetism is still not very clear. Earlier, Zaidi et al.[11] and Nascimento et al.[12] also reported ferromagnetic ordering at room temperature in different polymer systems and they concluded that magnetic impurities have no role to play in such magnetic ordering. They suggested that spin-½ polarons interact either ferromagnetically or anti-ferromagnetically depending on the polymer morphology to determine the magnetic state of the sample. In the polymeric diodes under present study, we believe that the injected carriers get trapped at the polymer defect sites and change the ground state magnetic ordering. Depending on the polymer morphology the nature of the trapped state can vary significantly and the carriers lying in the deep trapped state can sustain their magnetic orientation long after the external field is withdrawn or reversed and can give rise to extremely long relaxation time in these devices. This effect can not be associated to any permanent degradation of the device as upon removal of the electric field the samples regain their original resistance state. It is to be noted that similar long spin relaxation time have earlier been observed in inorganic ferromagnetic metal oxide and manganite materials which also gives rise to asymmetric magnetic hysteresis behavior. [13-15] In oxide semiconductor and half metal systems the relaxation strongly depends on oxygen vacancy and other defect sites which act as a trapping center for the carriers. The spin states of the charge carriers in the deep trapped sites flip after a longer time period compared to the free ones when the magnetic field direction is reversed. [13–15]

In summary, we have observed dependence of magnetoresistance of RRP3HT based diodes on the magnetic history of the device. Faster magnetic field sweep gives rise to broader line shape and decreased OMAR value while slower scan speed gives rise to increased OMAR and narrower line shapes suggesting charge carrier trapping plays a major role in OMAR. In light of the recently observed weak ferromagnetism in the polymers [10], we suggest that trapped carriers in presence of magnetic fields can give rise to magnetic clusters within the device with long spin relaxation time causing the hysteretic behavior of OMAR. These results also underline importance of necessary precautions for the OMAR measurements and the interpretation of the data.

The authors gratefully acknowledge the Wihuri Foundation and financial support from the Academy of Finland projects 116995 and 107684 through the Centre of Excellence Programme. Planar International Ltd. is acknowledged for the patterned ITO substrates.

## References

[1] T. L. Francis, Ö. Mermer, G. Veeraraghavan and M. Wohlgenannt, New J. Phys. **6**, 185 (2004).

[2] Y. Sheng, T. D. Nyugen, G. Veeraraghavan, Ö. Mermer, M. Wohlgenannt, S. Qiu, and U. Scherf, Phys. Rev. B **74**, 045213 (2006).

[3] E. L. Frankevich, A. A. Lymarev, I. Sokolik, F. E. Karasz, S. Blumstengel, R. Baughman and H. H. Hörhold, Phys. Rev. B **46**, 9320 (1992).

[4] V. Prigodin, J. Bergeson, D. Lincoln, and A. Epstein, Synth. Met. **156**, 757 (2006).

[5] P. Desai, P. Shakya, T. Kreouzis, W. P. Gillin, N. A. Morley, and M. R. J. Gibbs, Phys. Rev. B **75**, 094423 (2007).

[6] P. A. Bobbert, T. D. Nguyen, F. W. A. van Oost. B. Koopmans, and M. Wohlgenannt, Phys. Rev. Lett **99**, 216801 (2007).

[7] S. Majumdar, H. S. Majumdar, H. Aarnio, D. Vendarzande, R. Laiho, R. Österbacka, Phys. Rev. B **79**, 201202(R) (2009); S. Majumdar, H. S. Majumdar, H. Aarnio and R. Österbacka, EMRS Spring Meeting 2008 Proceedings (Springer) (in press);

[8] F. J. Wang, H. Bassler and Z. V. Vardeny, Phys. Rev. Lett. **101**, 236805 (2008).

[9] S. Majumdar, H. S. Majumdar, D. Töbjork and R. Österbacka, Phys. Stat. Solidi A (in press, arXiv: 0809.3864).

[10] S. Majumdar, H. S. Majumdar, J. O. Lill, J. Rajander, R. Laiho and R. Österbacka, Synth. Met. (submitted, arXiv:0905.1227).

[11] N. A. Zaidi, S. R. Giblin, I. Terry and A. P. Monkman, Polymer **45**, 5683 (2004).

[12] O. R. Nascimento, A. J. A. de Oliveira, E. C. Pereira, A. A. Correa and L. Walmsley, J. Phys.: Condens. Matter **20**, 035214 (2008).

[13] H. Huhtinen, R. Laiho and V. Zakhvalinskii, Phys. Rev. B **71**, 132404 (2005).

[14] T. Suominen, H. Huhtinen, S. Majumdar, P. Paturi, V. Zakhvalinskii and R. Laiho, J. Phys.: Condens. Matter (submitted).

[15] A. C. Pineda and S. P. Karna, J. Phys. Chem. A **104**, 4699 (2000).

**Figure captions:**

**Figure 1. (a)** % MR as a function of magnetic field for a typical RRP3HT diode measured with a bias current of 1 µA measured with different magnetic field scan speeds showing a huge change in OMAR value with longer exposure to magnetic field. **(b)** Line shape fit of the same OMAR curves showing power law fit of the OMAR value at higher magnetic fields for different magnetic field scan speeds. **(c)** % change in device resistance measured for a typical device while scanning magnetic field with $s = 20$ µT/sec and the device resistance under zero magnetic field on the same time scale.

**Figure 2. (a)** % MR as a function of magnetic field plot measured with same current of 10 µA and same scan speed with varying the intermediate time ($t_d$) between two successive scans. It is clearly seen that upon keeping the sample without electric and magnetic field for almost 10 minutes, the original resistance state of the sample is regained. **(b)** The device resistance as a function of time when first the magnetic field is scanned from 0 to 50 mT and then switched off. Simultaneous withdrawal of electric and magnetic field (red line) causes restoration of pristine device state within 5-7 minutes while only withdrawal of magnetic field does not allow full restoration of device resistance state (complete relaxation) even in hours.

**Figure 3.** The OMAR hysteresis loop for a typical RRP3HT diode with the scan speed of (a) 10 mT/sec and (b) 20 µT/sec showing the distinct time dependence of OMAR scans and closed loops only for the slowest scan.

**Figures:**

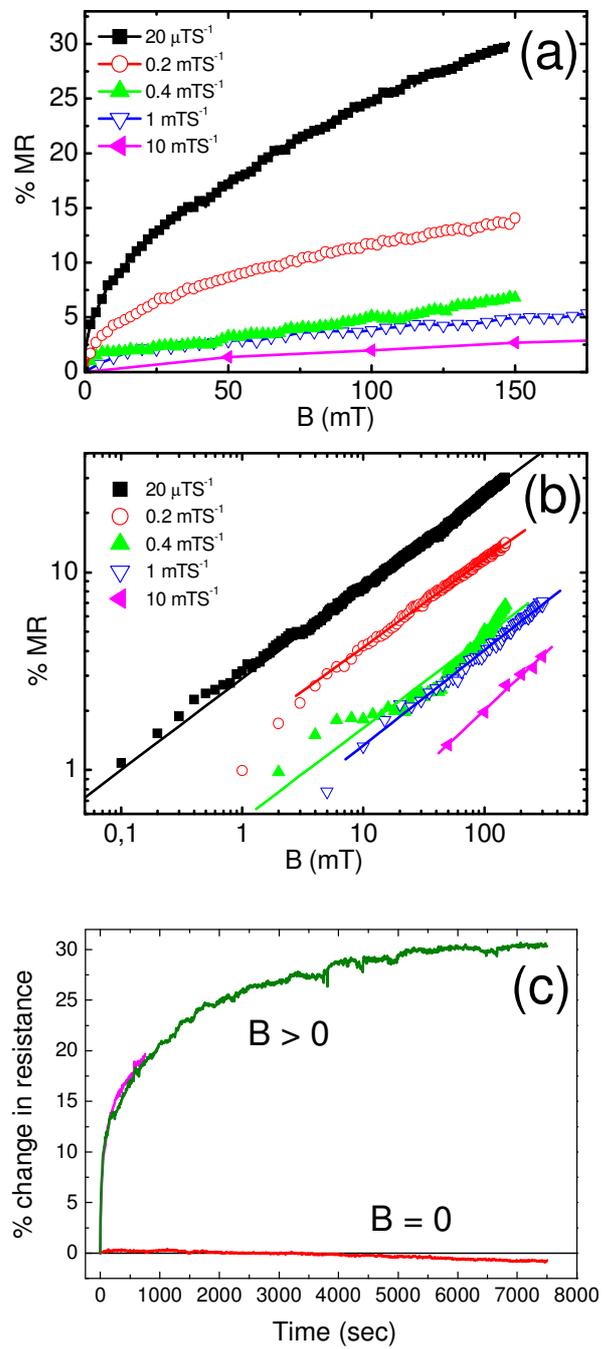

Figure 1. Sayani Majumdar et. al

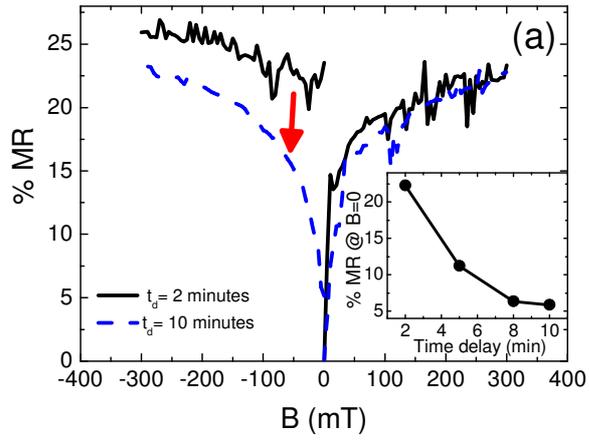

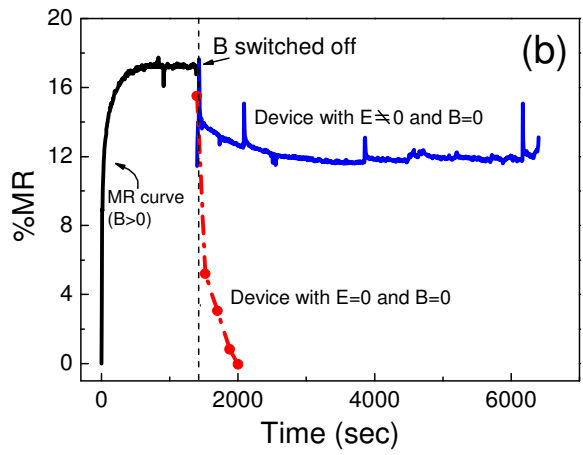

Figure 2. Sayani Majumdar et. al.

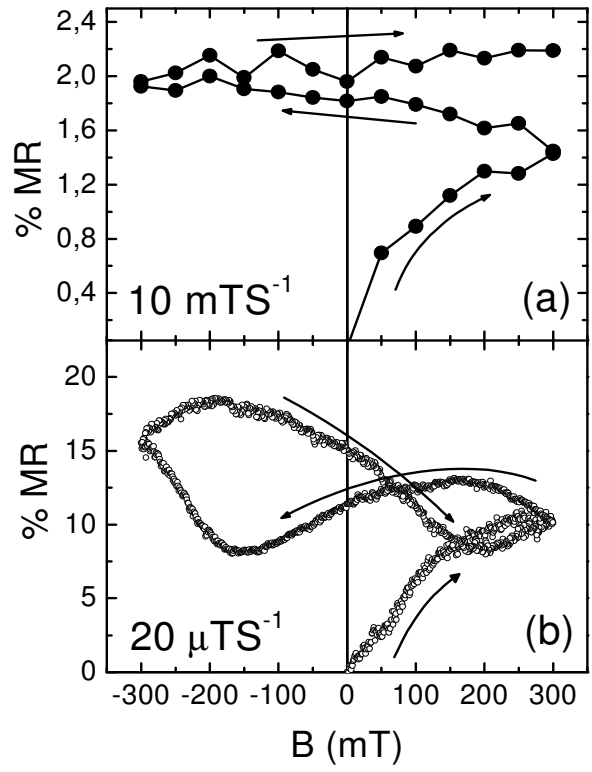

Figure 3. Sayani Majumdar et. al.